\documentclass[12pt]{iopart}
\usepackage{graphicx}
\usepackage{amssymb}
\usepackage{color}
\usepackage{booktabs}

\begin{document}
 
\title[Spectroscopic signatures of superconducting metal-decorated
graphene]{Spectroscopic signatures of different symmetries of the
  superconducting order parameter in metal-decorated graphene}

\author{Timo Saari$^{1}$, Jouko Nieminen$^{1,2}$ and Arun Bansil$^{2}$} 
\address{$^{1}$ Department of Physics, Tampere
  University of Technology, P.O. Box 692, FIN-33101 Tampere, Finland}
\address{$^{2}$ Department of
  Physics, Northeastern University, Boston, Massachusetts, USA}
\eads{\mailto{jouko.nieminen@tut.fi}}

\date{Version of \today}
\begin{abstract}

  Motivated by the recent experiments indicating superconductivity in
  metal-decorated graphene sheets, we investigate their quasi-particle
  structure within the framework of an effective tight-binding
  Hamiltonian augmented by appropriate BCS-like pairing terms for
  p-type order parameter. The normal state band structure of graphene
  is modified not only through interaction with adsorbed metal atoms,
  but also due to the folding of bands at Brillouin zone boundaries
  resulting from a $\sqrt{3}\times\sqrt{3}R30^{\circ}$
  reconstruction. Several different types of pairing symmetries are
  analyzed utilizing Nambu-Gorkov Green's function techniques to show
  that $p+ip$-symmetric nearest-neighbor pairing yields the
  most enhanced superconducting gap. The character of the order
  parameter depends on the nature of the atomic orbitals involved in
  the pairing process and exhibits interesting angular and radial
  asymmetries.  Finally, we suggest a method to distinguish between
  singlet and triplet type superconductivity in the presence of
  magnetic substitutional impurities using scanning tunneling
  spectroscopy.
\end{abstract}

\date{Version of \today} 


\maketitle

\section{Introduction}

Since its discovery, graphene has fascinated the scientific community
with its remarkable electronic properties, such as high electron
mobility and the anomalous quantum Hall
effect.~\cite{castroneto,Banszerus,Ostrovsky} Although pristine
graphene seems to lack superconductivity (SC), it can be induced via
the proximity effect\cite{Heersche}. More notably, SC state has been
found in intercalated graphite structures, especially CaC$_{6}$, where
metal atoms reside in the space between the loosely interacting
graphene
layers~\cite{weller,yang,bergeal,chapman,ludbrook,profeta}. Intriguingly,
from the point of view of their electronic structure, intercalated
graphite structures have also provided a promising platform for
developing high capacity rechargeable batteries \cite{yoo}.  The
findings of SC in these materials have motivated recent experiments,
which indicate that metal decoration might also induce SC in a single
graphene sheet~\cite{ludbrook,chapman}, although the nature of SC
remains unclear.

An early theoretical study of metal-decorated graphene by Uchoa and
Castro Neto \cite{uchoa} considered various pairing symmetries in the
presence of band folding effects of a
$\sqrt{3}\times\sqrt{3}R30^{\circ}$ reconstruction. That study
discusses electron-phonon and electron-plasmon mediated SC, and
suggests that the extended s-wave or p+ip -wave
  pairing with nearest neighbor matrix elements is more feasible than
  s-wave pairing with onsite matrix elements. Although
Ref.~\cite{uchoa} emphasizes the electron-plasmon mechanism, the
possibility of phonon mediated SC has attracted attention in
intercalated graphene~\cite{profeta} where \textit{ab initio}
electron-phonon coupling computations rule out multigap SC, but
support anisotropic pairing between electrons~\cite{sanna}.

\begin{figure}
  \includegraphics[width=1.0\textwidth]{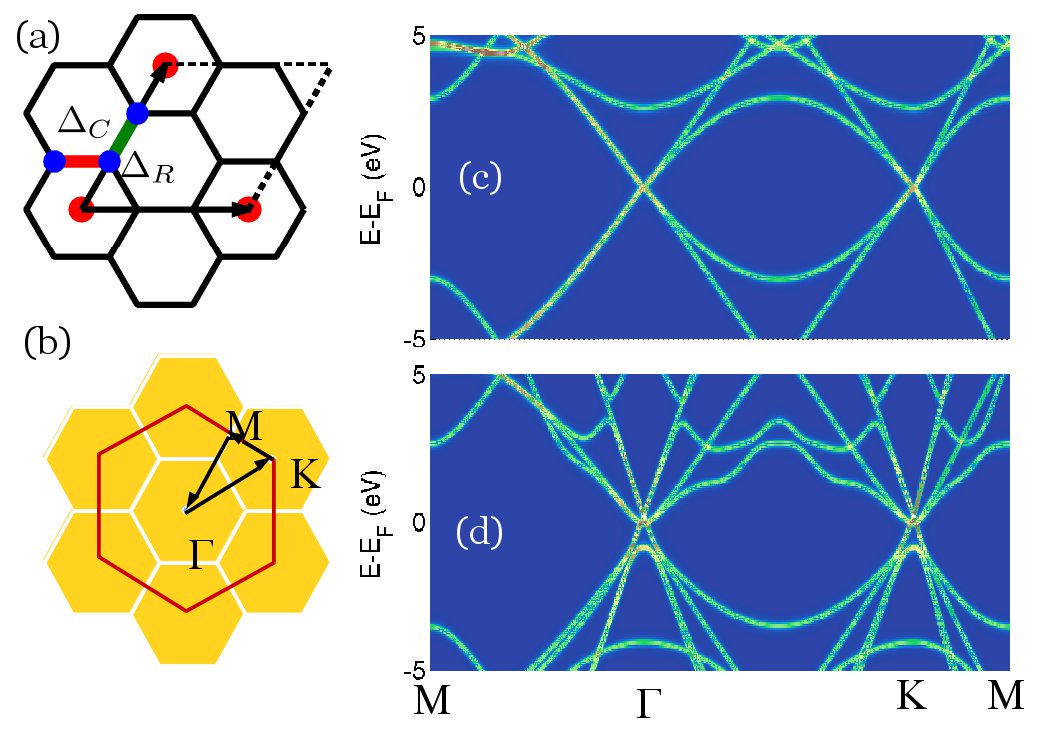}
  \caption{ (Color online) (a) Geometry of the $\sqrt{3}\times
    \sqrt{3} R 30^{\circ }$ structure of CaC$_{6}.$ The primitive cell
    is indicated as a parallellepiped. The range and direction of the
    anomalous Hamiltonian matrix elements $\Delta_{R}$ and
    $\Delta_{C}$ are indicated as green and red lines between nearest
    neighbor sites, respectively. (b) Reciprocal space of the
    $\sqrt{3}\times \sqrt{3} R 30^{\circ }$ structure, along with the
    first Brillouin zone of pristine graphene (red hexagon). (c)
    Folded band structure of pristine graphene along
    high symmetry directions marked in (b). (d) Folded
      band structure of metal-decorated graphene. }
\label{normal_0}
\end{figure}

The purpose of this study is to examine spectroscopic signatures of
different symmetries of the superconducting order parameter (OP) in
metal-decorated graphene. We take CaC$_6$ as an exemplar system, and
focus on the quasiparticle (QP) and scanning tunneling spectra (STS)
associated with specific OPs. We do not attempt to assess the nature
of the mechanism mediating pairing, but rather seek to unfold the
fingerprints of different symmetries of OPs in QP dispersions and the
related local densities of states. While the emphasis
  is on variations of $s$- and $p+ip$-symmetric singlet
  superconductivity, we also distinguish between singlet- and
  triplet-type pairing by introducing a magnetic impurity into the
  system. Our analysis is carried out within the framework of an
effective tight-binding (TB) Hamiltonian, which we augment with
appropriate pairing matrix elements to model various OPs. The
Hamiltonian is fitted to DFT calculations in order to correctly
capture the low-energy states and their orbital
characters. The realistic gap widths are of the order
  of $6-11meV$ (see, e.g., \cite{weller}), but we exaggerate the
  amplitudes of the anomalous terms in order to highlight pairing
  effects on the electronic structure. This allows us to focus on the
behavior of the salient consequences of different superconducting
order parameters.

\section{Methodology}

Our model Hamiltonian involves one $s$-orbital and
three $p$-orbitals for each atom. The electron, hole and spin degrees
of freedom are incorporated as follows:
\begin{equation}
\fl \hat{H} = \sum_{\alpha\beta\sigma}
(\varepsilon_{\alpha}c^{\dagger}_{\alpha \sigma} c_{\alpha \sigma}+
V_{\alpha \beta}
c^{\dagger}_{\alpha \sigma} c_{\beta\sigma}
 )+\hat{H}_{\textrm{SC}}+\hat{H}_{\textrm{MAG}}.
\label{hamiltonian}
\end{equation}
Here $c^{\dagger}_{\alpha \sigma}$ ($c_{\alpha \sigma}$) is the
real-space creation (annihilation) operator, $\alpha$ is a composite
index which encodes both the site and orbital information, and
$\sigma$ is the spin index. The on-site orbital energy
($\varepsilon_\alpha$) and the hopping integral between orbitals
$\alpha$ and $\beta$ ($V_{\alpha\beta}$) are obtained within the
Slater-Koster formalism~\cite{SK,Harrison}. All parameters in the
normal state part of the TB Hamiltonian are fitted to the low-energy
DFT band structure of CaC$_{6}$ obtained using the Quantum
Espresso~\cite{QE,pseudo} package. Fig. 1(a) and
  1(b) depict the real space structure of the system indicating the
  structures for pristine honeycomb lattice and the lattice with the
  reduced symmetry. Fig. 1(c) Shows the folded band structure for
  the pristine graphene and, for comparison, 1(d) shows the effect of
  Ca decoration.  A more detailed analysis of the band structure is
  given in Appendix A.

For the SC part of the Hamiltonian, $\hat{H}_{SC}$, different OPs
(anomalous matrix elements of the Hamiltonian) $\Delta_{\alpha\sigma
  \beta \sigma'}(\lambda)$ are modeled through the choices of atomic
orbitals, spin degrees of freedom, and the pairing symmetries (labeled
$\lambda$, see also Appendix B). However, in all cases, we have
artificially enhanced the amplitudes of OP matrix elements to better
identify and highlight the gaps in QP dispersions and local densities
of states (LDOSs). Thus, we write the SC part of the Hamiltonian as
\begin{equation}
\fl \hat{H}_{SC} = \sum_{\alpha\beta\sigma}
(\Delta_{\alpha \sigma \beta \sigma'}(\lambda) c^{\dagger}_{\alpha \sigma}
c^{\dagger}_{\beta -\sigma} 
+  \Delta_{\beta \sigma' \alpha \sigma}^{\dagger}(\lambda)
 c_{\beta -\sigma} c_{\alpha \sigma} 
 ).
\label{hamiltonian_SC}
\end{equation}
It is important to distinguish between the symmetry of the order
parameter $\lambda$ and the character of the involved atomic orbitals
$\alpha$ and $\beta$.  For example, if the matrix element
$\Delta_{\alpha \beta}(\lambda)$ has $p+ip$-symmetry,
its complex phase is the same as the phase of $x_{\alpha
  \beta}+iy_{\alpha \beta}$, where $x$ and $y$ refer to the relative
coordinates of the atoms with orbitals $\alpha$ and $\beta.$
To put it simple, we mainly use the symmetry choices
  of Ref. \cite{uchoa} where, in addition to spherically symmetric
  onsite matrix elements of s-wave order parameter, there are also
  nearest neighbour matrix elements which can expanded to follow
  $\propto k_{x}+ik_{y}$ (see also
  Refs. \cite{doniach1,doniach2}). The main novelty here is that in
  Ref. \cite{uchoa} the basis consists of $p_{z}$ orbitals of carbon,
  but in our cases the basis is significantly larger (See also
  appendices A and C).

We will find that the QP-dispersion of SC singlet and triplet pairings are
indistinguishable unless a spin-dependent perturbation is
present. With this in mind, we allow the possibility of introducing a
substitutional magnetic impurity into the Hamiltonian (1) via the term
\begin{equation}
\hat{H}_{MAG} = Um_{\gamma}(c^{\dagger}_{\gamma \uparrow} c_{\gamma \uparrow}-
c^{\dagger}_{\gamma \downarrow} c_{\gamma \downarrow}).
\label{hamiltonian_MAG}
\end{equation}
Here $\gamma$ refers to the index of the orbital contributing to local
magnetic moment.  The impurity is modeled by replacing one metal atom with a
model atom (see Fig. \ref{magsts}(a) insert), where the two spin
states are split via differences in their on-site energies. In order
to create a visible effect, we have taken $Um_{\gamma} = \pm 1.0eV$
for the spin-up and spin-down $p$-orbitals of the impurity atom,
respectively.

We analyze the electronic structure generated by the Hamiltonian by
utilizing Bogoliubov-de Gennes equations and the associated tensor
(Nambu-Gorkov) Green's function ${\cal G}$\cite{Abrikosov,Fetter}:
\begin{equation}
  {\cal G} =  {\cal G}^0 +  {\cal G}  \bold{\cal D} {\cal G}^0
\label{dyson1}, 
\end{equation}
where ${\cal G}^{0}$ is the Nambu-Gorkov Green's function without
electron-hole interaction,
\begin{displaymath}
  {\cal G} =
\left(
   \begin{array}{cc}
G_{e}& F\\
F^{\dagger}& G_{h}
   \end{array}
\right)~\textrm{with}~
{\boldmath c}_{\alpha} =
\left(
  \begin{array}{c}
c_{\alpha \uparrow}\\
c_{\alpha \downarrow}\\
c^{\dagger}_{\alpha \uparrow}\\
c^{\dagger}_{\alpha \downarrow}
  \end{array}
\right)
\end{displaymath}
and
\begin{displaymath}
  \bold{\cal D} =
\left(
   \begin{array}{cc}
0& \tau\\
\tau^{\dagger}& 0
   \end{array}
\right)
\end{displaymath}
Here, $G_{e}$ and $G_{h}$ denote the spin-resolved Green's functions
for electrons and holes, respectively\footnote{Spin-flip terms are
  neglected in the present calculations.}, and the matrix elements of
the operator $\tau$ represent the interaction terms of the Hamiltonian
of Eq.~\ref{hamiltonian}. As in Refs. \cite{paulsson, nieminenPRB},
the elements of Nambu-Gorkov Green's function provide us the local
density of states
\begin{equation}
	\rho_{\alpha\sigma, \beta \sigma'} = 
-\frac{1}{2\pi i}\left(\cal G^{+}_{\alpha\sigma, \beta\sigma'}-
\cal G^{-}_{\beta\sigma',\alpha\sigma}\right)
\end{equation}
and the electron-hole pairing amplitude is
\begin{equation}
	\rho_{eh} = \textrm{Tr}(FF^\dagger).
\end{equation}
The preceding equations allow us to obtain contributions to various
quantities from different orbitals, as well as from the electron,
hole, or spin degrees of freedom. The foremost use of
  the density matrix is different presentations of energy states as a
  function of different degrees of freedom. For example the
  QP dispersion can be expressed as $\rho(E,k)$-diagram,
  which is essentially the band diagram. Furthermore, one can take a
  trace of $\rho(E,k)$ over the electron part of the basis as is done
  in most of the QP dispersions presented in this work, or to
consider the anomalous electron-hole terms as in Fig. 3. (c-f) 
(see also Ref. \cite{nieminenPRB}).

Another use of the density matrix is simulations of
  scanning tunneling microscopy/spectroscopy (STM/STS), where we
apply the Todorov-Pendry approach~\cite{Todorov} (see also Ref.
\cite{Tersoff}) in which the differential conductance $\sigma$ between
orbitals of the tip ($t,t'$) and the sample ($s,s'$) is given
by\cite{NLMB,nieminenPRB}
\begin{equation}
\sigma = \frac{dI}{dV} = \frac{2 \pi e^2 }{ \hbar} \sum_{t t' s s'}
\rho_{tt'}(E_F)V_{t's} \rho_{ss'}^{}(E_F+eV)V_{s't}^{\dagger}.
\label{conductance}
\end{equation}

We will see that structural variations in the SC gap do not lead to
variations in the LDOS, which are pronounced enough to allow
identification of the underlying coupling mechanism via regular
dI/dV-spectroscopy. We consider therefore the effect of a local
magnetic impurity to determine how this perturbation will be seen in
STM/STS under various pairing symmetries. In addition to the regular
STM topographic maps, we compute current polarization maps in constant
current mode where the regular dI/dV spectrum is scaled by the normal
state spectrum and the polarized differential conductance spectrum,
Here, current polarization, following Ref. \cite{wortmann}, is
defined as:
\begin{equation}
  P_{I} = \frac{I_{\uparrow}-I_{\downarrow}}{I_{\uparrow}+I_{\downarrow}}
\label{currentpol}
\end{equation}
where current can be obtained with numerical integration
of equation \ref{conductance}. Still following \cite{wortmann},
the differential conductance polarization is
\begin{equation}
  P_{\sigma} = \frac{\frac{d I}{dV}_{\uparrow}-\frac{d I}{dV}_{\downarrow}}{\frac{d I}{dV}_{\uparrow}+\frac{d I}{dV}_{\downarrow}}.
\label{diffpol}
\end{equation}
Note that these expressions refer to the case where components of
spin are perpendicular to the sample surface, although practical
spin-resolved STM often involves filtering components
parallel to the surface \cite{wortmann, wiesendanger}. The modeling of
the more complicated case of spin-filtering in the presence of
horizontal spin components will be considered elsewhere.

\section{Results} 

The following discussion will emphasize three main points: (1) Since
the  singlet p+ip-wave symmetry of the OP
$\Delta(\lambda)$ seems to be the most effective route toward forming
the SC gap, we will focus on different orbital combinations for
creating a p+ip-wave OP. (2) Concerning the normal-state band
structure, the metal atoms lead to a Kekul\'e-type folding
\cite{Gomes,Paavilainen} of the $\pi^{*}$-band with a gap and, in
addition, new conical non-folded bands appear due to rehybridization
of the p-orbitals of carbon and metal atoms, see Appendix A for
details.  To determine how these folded and conical bands contribute
to SC and the OP, we consider OP matrix elements between the relevant
atomic orbitals in two scenarios: (a) Use $p_{z}$ orbitals of
neighboring carbon atoms to construct $\pi$-type OP or (b) use
horizontal p-orbitals of carbon to construct $\sigma$- or $\pi$-type
OP. In this connection, we discuss the possibility of non-equivalent
$\Delta_{R}$ (radial) and $\Delta_{C}$ (angular)
terms.  (3) Finally, although we focus on singlet $p+ip$
  superconductivity, we consider the possibility to experimentally
  distinguish between singlet and triplet cases in the presence of
  magnetic impurity.

\subsection{SC state: $\sigma$ vs. $\pi$-type order parameter}

\begin{figure}
  \includegraphics[width=1.0\textwidth]{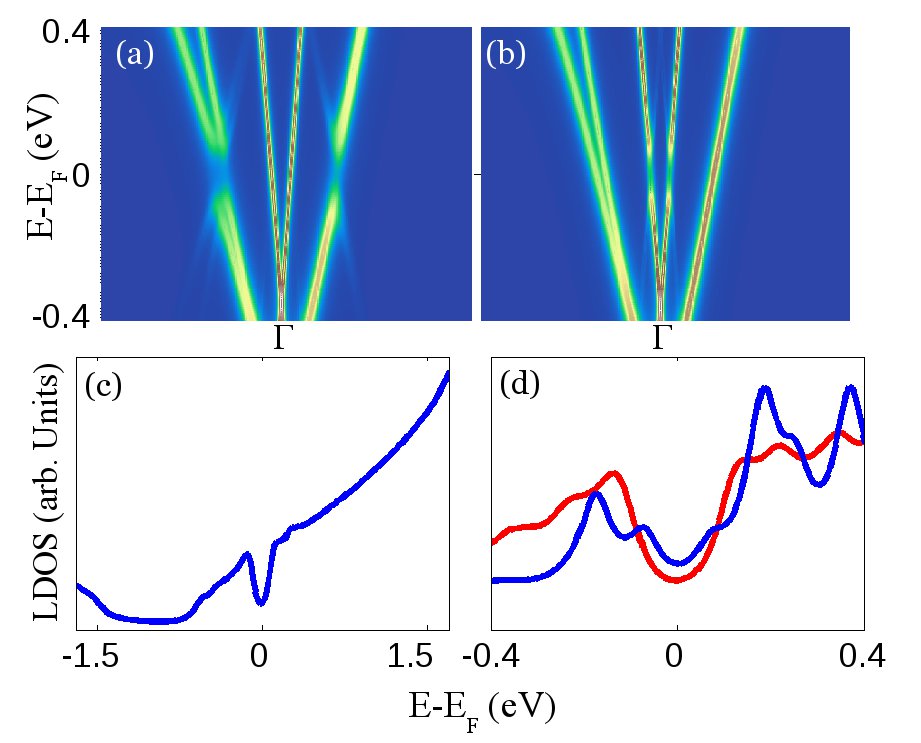}
  \caption{(Color online) QP dispersion for electron doped
    CaC$_{6}$ when the matrix elements of $\Delta$ are between (a)
    $p_z$-orbitals and (b) $p_{x/y}$-orbitals of neighbouring C atoms.
    (c) Contribution from $p_z$(C) to the density of states
    corresponding to the parabolic bands in (a). (d)
    PDOS from p$_{z}$(C) (red) and p$_{x}$(C)+p$_{y}$(C) in the region
    near $E_F$ (blue line is multiplied by a factor of 10).}
\label{DOS}
\end{figure}

We start by considering the effect on SC of different types of OPs
with $p_{x}+ip_{y}$ symmetry in terms of contributing atomic orbitals.
As the first case, we scrutinize Hamiltonian matrix elements
$\Delta_{\alpha\beta}(\lambda)$, where $\alpha$ and $\beta$ are the
$p_{z}$ orbitals of the two neighboring carbon atoms. This can be
characterized as p-type coupling with a $\pi$-type orbital-orbital
character. Such a matrix element opens up a gap with coherence peaks
for the $\pi^{*}$ bands under electron or hole doping, see
Fig.~\ref{DOS}(a), which is similar to the observations in
intercalated bulk CaC$_6$ in Ref.~\cite{bergeal}. Since these
matrix elements directly couple only with the $p_{z}(C)$-orbitals, they
have little effect on the conical bands and, as a result, they remain
essentially intact.

These interlayer bands (IL) (see Appendix A) with a
Dirac point mainly involve $p_{x/y}$(C) character, and therefore, we
must also consider the p-wave matrix elements $\Delta_{\alpha
  \beta}(\lambda)$, where $\alpha$ and $\beta$ are linear combinations
of the $p_{x}$ and $p_{y}$ orbitals of two neighbouring carbon
atoms. There are two possibilities: the orbitals can be combined to
make a $\sigma$-type combination, where the hybridized p-orbitals
point along the bond between the two carbon atoms or a $\pi$-type
combination, where the orbital is oriented in the perpendicular
direction.  Both these $\pi$ and $\sigma$-type matrix elements open up
an SC gap uniformly at $\Gamma$ when the Fermi-energy lies outside the
gap of the folded $\pi^{*}$ band, see Fig.~\ref{DOS}(b); when $E_F$
lies within the gap, the SC gap opens only due to the $\sigma$- but
not the $\pi$-type matrix element.  This directional dependence is
likely dependent on the hybridization of the horizontal $p$-orbitals
with $p_z$(C)-orbitals, which contribute to the conical bands only
within the gap of the $\pi^{*}$ band (See discussion of orbital contributions in Appendix A). 
\footnote{ We have obtained the
  corresponding spectra for triplet pairing~\cite{Qi}, but found no
  difference from the singlet case. This is to be expected, since the
  Hamiltonian contains no spin-orbit coupling or magnetic
  order. $s$-wave singlet pairing with an on-site matrix element can
  also lead to pairing around $\Gamma$, but none of these pairing
  types affect the IL-bands.}


The distinction between the two kinds of orbital contributions can
also be seen in Fig.~\ref{DOS}(c), where an inspection of the $p_z$(C)
contribution to the partial density of states (PDOS) shows the
presence of a clear SC gap.  If the horizontal p-orbital contribution
is taken into account as well, the PDOS projected onto $p_{x/y}$(C)
also shows an SC gap (see Fig.~\ref{DOS}(d)), but the intensity is an
order of magnitude smaller.

\subsection{Anisotropy of order parameter $\Delta$: radial vs. angular bonds}

In constructing SC matrix elements, it is useful to make a distinction
between the \emph{radial} bonds [$\Delta_R$, Fig. 1(a)], which connect the two
carbon atoms between the neighboring metal atoms, and what may be called
\emph{angular} bonds [$\Delta_C$, Fig. 1(a)], which connect
phenyl-ring-like hexagons around the metal atoms.

\begin{figure}
  \includegraphics[width=1.0\textwidth]{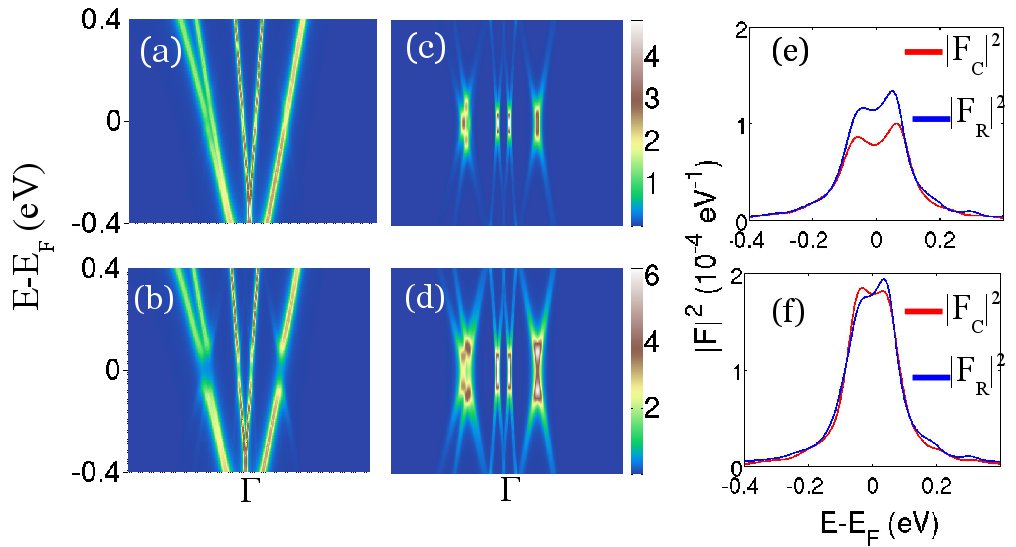}
  \caption{(Color online) QP dispersion for electron doped
    CaC$_{6}$ when only (a) the radial and (b) the angular matrix elements
    of $\Delta$ are included for all
    $p$-orbitals.  
     (c) and (d) Energy-momentum dispersion of
      electron-hole pairing amplitudes
      corresponding to (a) and (b), respectively.
      (e) and (f) Matrix elements $\vert F_{R} \vert^{2}$ (blue curves) and $\vert F_{C} \vert^{2}$ (red curves) of electron-hole pairing
      amplitudes, where (e) and (f) correspond to (a) and (b),
      respectively.  }
\label{fig3}
\end{figure}

Our calculations indicate that the angular matrix element,
$\Delta_{C}$, contributes strongly to gap formation,
see Figs. \ref{fig3} (a) and (b), and that the radial
matrix element alone is not sufficient for opening an SC gap in the
electronic spectrum.  The E-k-dispersion in
{Fig. \ref{fig3}} further indicates that the
{anomalous amplitudes} $\vert F \vert^{2}$ lead to
some QP-hybridization via both types of matrix elements, but the
angular symmetry plays a dominant role. Interestingly, however, there
is little difference between the amplitudes of the resulting outgoing
radial and angular matrix elements of the anomalous Green's function,
$\vert F_{C} \vert^{2}$ and $\vert F_{R} \vert^{2}.$ Since 
the OP $\Delta$ would be coupled self-consistently with $F$
\cite{Abrikosov, Fetter}, the directional homogeneity of the anomalous
Green's function would indicate connections between the symmetry of
the OP and the bosonic mechanism underlying SC. In particular, a
directionally anisotropic OP can only be obtained if the related
bosonic modes are directionally anisotropic.

\subsection{STM/STS and pairing mechanism in the presence of a
magnetic impurity: singlet vs. triplet pairing}

\begin{figure}
  \includegraphics[width=1.0\textwidth]{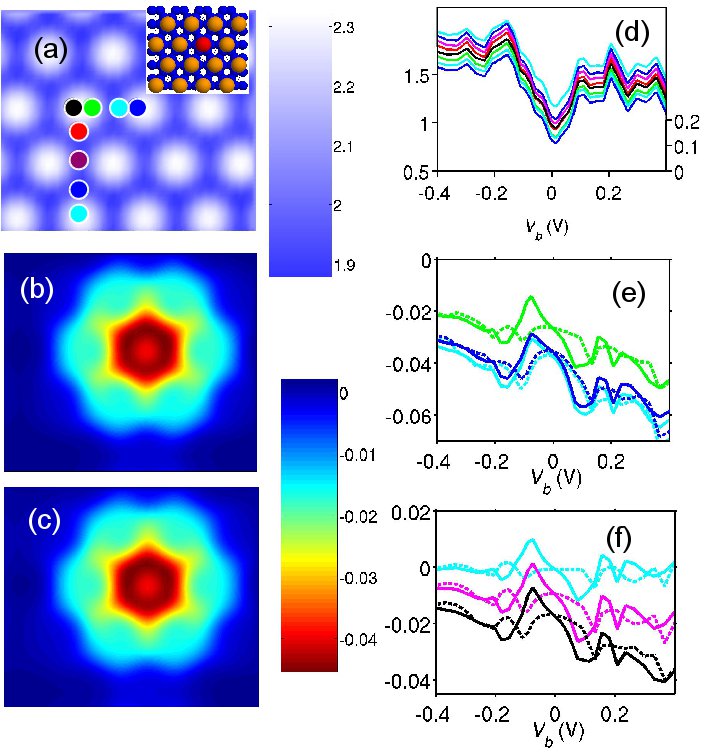}
  \caption{ (Color online) (a) Simulated STM topographic map for
    CaC$_{6}$ in the presence of a magnetic impurity. Colored dots
    indicate the real space positions where the corresponding computed
    STS spectra are shown in (d). The inset shows the simulated
    primitive cell with C (blue) and Ca (yellow) atoms along with the
    impurity atom (red). (b) and (c) Spin-polarized topographic map
    for the system in (a) for singlet- and triplet-type pairing,
    respectively. (d) dI/dV spectra for the SC state scaled by the
    normal state spectra at positions indicated in (a). Lines with
    various colors correspond to tip positions given by the dots of
    the same color in (a). All spectra have been scaled by the
    corresponding normal state spectra. (e) Spin-polarized
    dI/dV-spectra computed at three different points along the
    horizontal line joining the impurity atom in the inset in (a) with
    Ca.  Results for singlet pairing (solid lines)
    and triplet pairing (dashed lines) are shown. (f)
    Spin-polarized dI/dV-spectra computed at three different points
    along the vertical line joining the two Ca atoms.  Results for
    singlet pairing (solid lines) and
    triplet pairing (dashed lines)} are shown.
\label{magsts}
\end{figure}

As noted already, we expect little difference between the STS/STM
spectra of CaC$_6$ for singlet and triplet pairing. However, since a
singlet Cooper pair has total spin $S=0$, and triplet has $S=1$, a
difference could be induced via a magnetic perturbation. Accordingly,
we consider the effects of substituting a Ca atom with a magnetic
impurity. Note that the STM corrugation map of Fig \ref{magsts}(a)
shows the metal atoms as bright spheres, and gives no hint of the
magnetic perturbation. Similarly, scaled dI/dV spectra in
Fig. \ref{magsts}(d) computed at various positions (see colored dots
in Fig.~\ref{magsts}(a)) show little variation with
position. Qualitatively, the same kind of scaled set of spectra are
obtained for both the singlet and the triplet case.  We thus adduce
that a regular STM/STS measurement will not detect the presence of a
magnetic impurity.

Fig. \ref{magsts}(b) shows that when we consider a map of polarized
current, $P_{I}$-map (Eq. \ref{currentpol}), the magnetic impurity at
the center of the figure is clearly detected, with the perturbation
extending essentially only to the neighboring Ca atoms.  The
polarization is seen to be the strongest on the six carbon atoms
surrounding the impurity, being nearly as strong as it is on the
impurity atom. A hexagonal pattern of slightly lower (ferromagnetic)
polarization is seen in the neighborhood of the six Ca atoms with the
magnetic effect rapidly dying out as we move further away from the
impurity. Note that this map is practically identical for both
singlet- and triplet-case (Fig. \ref{magsts}(b) and (c)). This
indicates that the $P_I$-map also cannot be used to distinguish
between singlet and triplet pairing, even though this map clearly
shows the magnetic perturbation.

Figures \ref{magsts}(e) and (f) finally consider polarized
differential conductance, $P_{\sigma}$, see Eq. (\ref{diffpol}). The
spectra are now seen to distinguish between singlet
  (solid lines) and triplet (dashed line) pairing around the magnetic
  impurity. For the singlet case, polarization changes abruptly around
  the coherence peaks and varies roughly linearly in the gap
  region. In sharp contrast, in the triplet case, we see a minimum at
  the coherence peak energies, and a maximum at energies between these
  peaks. These fingerprints of singlet and triplet pairing should be
  observable in spin-polarized dI/dV-spectra, and allow thus a handle
  on the underlying pairing mechanism.

\section{Summary and Conclusions}

We consider p+ip -wave singlet superconductivity in
metal-decorated graphene within the framework of a tight-binding
Hamiltonian based on first-principles normal state band structure, and
discuss the characteristic spectroscopic fingerprints of different
superconducting order parameters.

Both the in-plane $p_{x/y}$(C)-orbitals and the out-of-the-plane
$p_z$(C)-orbitals are needed to open up a superconducting gap.
Anomalous matrix elements between the $p_z$-orbitals open a gap
between the $\pi^*$-bands whereas matrix elements between the
$p_{x/y}$-orbitals are required to open the gap in the conical
interlayer bands.  Therefore, ARPES experiments with sufficiently high
resolution could distinguish between superconducting gaps in different
bands, and could thus be used to identify the atomic orbitals involved
in the underlying pairing mechanism. On the other hand, although a few
meV gap expected in intercalated graphite \cite{weller} could be
observed via STM/STS experiments as indicated by the results of
Fig.~\ref{DOS}(c), one would not be able to distinguish between
different pairing symmetries from the measured spectra.

Due to the Kekul\'e structure induced by metal decoration, the order
parameter is anisotropic. As a result, the superconducting gap forms
mainly due to the angular matrix elements, which reflect the couplings
between the neighboring carbons circling a metal atom.  Unfortunately,
there is no direct way to experimentally detect directional anisotropy
in the order parameter.  One could speculate about the possibility of
obtaining the anomalous QP spectrum through a measurement using a
superconducting STM tip, where the directional anisotropy might be
reflected in the quasiparticle interference (QPI) patterns.

Our analysis shows that the character of the SC gap depends on the
nature of the atomic orbitals at the Fermi energy involved in the
pairing process, which drive interesting angular and radial
asymmetries in the SC order parameter.  The computed STM/STS spectra
with and without a magnetic impurity indicate that a magnetic impurity
will essentially be invisible in a standard (spin-unresolved)
spectrum. This, however, is not the case in a spin-resolved STM/STS
spectrum, where the polarization around the impurity can be seen
clearly, and singlet vs. triplet pairing can be distinguished in the
polarized differential conductance spectrum of Eq. \ref{diffpol}. Our
study indicates that spin-polarized measurements would provide new
insight into the nature of the order parameter and its symmetry in
metal-decorated graphene systems. An interesting prospect is, if
superconductivity of graphene could be tuned with modulations in metal
decoration. Based on the calculated effects of magnetic impurity, as
well as the dependence of the order parameter on the folded bands
vs. decoration induced conical band, we suggest STM/STS experiments on
metal decoration with magnetic and non-magnetic substitutional
impurities.

{\bf Acknowledgments} This work benefited from the resources of
Institute of Advanced Computing, Tampere. T.S. is grateful to
V\"ais\"al\"a Foundation for financial support. The work at
Northeastern University was supported by the US Department of Energy
(DOE), Office of Science, Basic Energy Sciences grant number
DE-FG02-07ER46352 (core research), and benefited from Northeastern
University's Advanced Scientific Computation Center (ASCC), the NERSC
supercomputing center through DOE grant number DE-AC02-05CH11231, and
support (applications to layered materials) from the DOE EFRC: Center
for the Computational Design of Functional Layered Materials (CCDM)
under DE-SC0012575. Conversations with Esa R\"as\"anen and Sami Paavilainen 
are gratefully acknowledged.

\appendix 

\section{Geometric and electronic structure and normal state band characters}

Figs. \ref{normal_0}(a) and (b) show how the $\sqrt{3}\times
\sqrt{3}R30^{\circ}$ reconstruction reduces lattice symmetry. As a
result, bands fold at the two inequivalent $K$-points of the large BZ
of pristine graphene to the $\Gamma$-point of the small BZ of the
metal-decorated graphene sheet. The new lattice of C atoms can also be
viewed as a \emph{Kekul\'e}-distorted graphene lattice (see Refs.
\cite{Gomes,Paavilainen}).  This distortion opens a gap at the
Dirac point of the $\pi^{*}$-type bands, which is clearly seen in the
band structures of Fig. \ref{normal_0}(c) and (d). As is well known
\cite{uchoa,profeta,yang}, in addition to the $\pi^{*}$-bands,
``hourglass''-like bands are formed from the $sp^{2}$-hybridized
horizontal $p$-orbitals of C atoms and orbitals of the metal atoms. We
refer to these two types of bands as $\pi^{*}$- and interlayer-bands
(IL-bands).
This nomenclature,
however, is not followed consistently in the literature, and for this
reason, we comment further on this point.

\begin{figure}
  \includegraphics[width=1.0\textwidth]{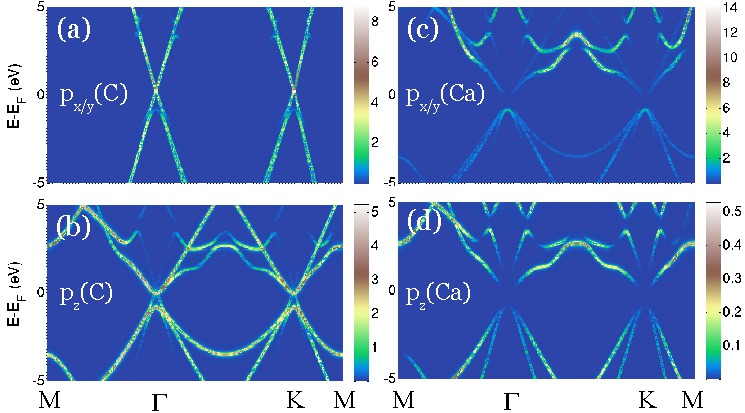}
  \caption{(Color online) Contribution of various orbitals in the band
    structure of Fig.~\ref{normal_0}(d) for decorated CaC$_{6}$: (a)
    Horizontal p-orbitals of C; (b) Folded $\pi^{*}$ orbitals of C;
    (c) Horizontal $p$-orbitals of Ca; (d) $p_{z}$-orbitals of Ca.}
\label{phasediagN}
\end{figure}

Since the gapped or \emph{Kekul\'e}-distorted bands are doubly
degenerate (in addition to spin degeneracy), they are folded from the
graphene $K$-points. The LDOS decomposition in Fig. \ref{phasediagN}
shows that these bands possess a strong p$_{z}$(C)-character
especially in the vicinity of the gap; the Ca orbitals mix with these
bands at higher energies. The conical IL-band, on the other hand,
merely possesses the spin-degeneracy and it is, therefore, a genuine
decoration-induced feature at the $\Gamma$-point. An analysis of the
wavefunction shows that IL-band is dominated by p$_{x/y}$(C) orbitals,
which originate from $sp^{2}$-hybridization. Since orbitals of Ca
atoms overlap weakly with the horizontal p(C)-orbitals,
Kekul\'e-distortion does not open a gap in these bands.

\section{Implementation of singlet and triplet superconductivity in
  the tight binding Hamiltonian}

Here we consider anomalous matrix elements of the Hamiltonian in using
the basis set: $(\vert \alpha \uparrow \rangle, \vert\beta \uparrow
\rangle, \vert \alpha \downarrow \rangle,\vert \beta\downarrow
\rangle).$ For a singlet configuration ($s=0$), antisymmetric
two-particle states with $m_{s}=0$ are of the form $\vert \alpha
\uparrow \beta\downarrow \rangle - \vert \alpha \downarrow
\beta\uparrow \rangle.$ For a triplet ($s=1$), the state is symmetric
with respect to spin flip, and hence $\vert \alpha \uparrow
\beta\downarrow \rangle + \vert \alpha \downarrow \beta\uparrow
\rangle$ corresponds to $m_{s}=1$, whereas $\vert \alpha \uparrow
\beta\uparrow \rangle$ and $\vert \alpha \downarrow \beta\downarrow
\rangle$ stands for cases $m_{s} = 1$ and $-1$,
respectively. Construction of the order parameters then follow the
derivation given in Refs. \cite{Qi,Bernevik} for topological
superconductors.

In constructing the order parameters, we assume
that the combined angular momentum is $J=0$, which couples 
orbital and spin quantum numbers as: $m_{l}+m_{s}=0$, i.e., $m_{l} =
-m_{s}.$ For the singlet state we need an s-wave order parameter
$\Delta(s)$ and a sub-Hamiltonian for the four orbitals given by:
  
\begin{displaymath}
  \left(
   \begin{array}{cccc}
0& 0 &0 &\Delta(s)\\
0& 0 &\Delta(s) &0 \\
0 &-\Delta(s) &0 & 0\\
-\Delta(s) &0 &0 & 0
   \end{array}
\right)
\end{displaymath}
since spin-flip changes the sign of the matrix element.

 In addition to the symmetry with respect to the
  orbital/spin permutations, one must also account for the directional
  dependence of the order parameter $\Delta.$ Define $x_{\alpha \beta}
  = x_{\alpha}-x_{\beta}$, and likewise for the other coordinates, the
  form apart from an amplitude prefactor is as follows [$\Delta(s)$ is
  just a complex number (totally symmetric)]:
\begin{displaymath}
\Delta(z) \propto \frac{z_{\alpha \beta}}{r_{\alpha \beta}} = 
\cos{(\theta_{\alpha \beta})} 
\end{displaymath}
and 
\begin{displaymath}
  \Delta(\pm) \propto \frac{(x_{\alpha \beta}\pm i y_{\alpha
    \beta} )}{r_{\alpha \beta}}
= \sin{(\theta_{\alpha \beta})}e^{(\pm i \varphi_{\alpha \beta})}. 
\end{displaymath}
Hence, this scaling takes into account the dimensionality of the
system by introducing the appropriate rotational angles.

In the triplet case, we need a spatially antisymmetric wave function,
i.e., we need p-symmetric matrix elements. If we first look at the
case $m_{s}=0$, we need a $p_{z}$-symmetric matrix element
$\Delta(z).$ Since switching the order of the orbitals in the
two-particle state changes the sign, the sub-Hamiltonian goes into the
form:
\begin{displaymath}
  \left(
   \begin{array}{cccc}
0& 0 &0 &\Delta(z)\\
0& 0 &-\Delta(z) &0 \\
0 &\Delta(z) &0 & 0\\
-\Delta(z) &0 &0 & 0
   \end{array}
\right).
\end{displaymath}

For $m_{s}= \pm 1$, the symmetry of the matrix element is
$p_{\mp} = p_{x}\mp ip_{y}.$ Following the anticommutation arguments
for matrix elements $\Delta(\mp)$, we end up with the following
sub-Hamiltonian:
\begin{displaymath}
  \left(
   \begin{array}{cccc}
0 &\Delta(-) & 0& 0  \\
-\Delta(-) &0& 0& 0  \\
 0& 0&0 &\Delta(+)    \\
 0& 0&-\Delta(+) &0   \\
   \end{array}
\right).
\end{displaymath}
Hence, these matrix elements consist of up-up and down-down terms.
Finally, the total electron-hole block will be:
\begin{displaymath}
  \left(
   \begin{array}{cccc}
0 &\Delta(-) & 0& \Delta(s)+\Delta(z)  \\
-\Delta(-) &0& \Delta(s)-\Delta(z)& 0  \\
 0& -\Delta(s)+\Delta(z)&0 &\Delta(+)    \\
 -\Delta(s)-\Delta(z)& 0&-\Delta(+) &0   \\
   \end{array}
\right).
\end{displaymath}

\section{Parametrization of the Hamiltonian.}

We utilize the Slater-Koster tables \cite{SK,Harrison} to determine
the distance and directional dependence of the Hamiltonian matrix
elements, but their amplitudes are fitted to QE calculated band
structures especially at low energies.
The amplitudes for the matrix elements and the onsite terms are as follows:
\begin{table}[h!]
  \centering
  \caption{Amplitudes of the Slater-Koster terms.}
  \label{tab:table1}
  \begin{tabular}{l||r}
    overlap & amplitude\\
    \hline
    $ss\sigma$ & -0.80 \\
    $sp\sigma$ & 0.24 \\
    $pp\sigma$ & 3.24 \\
    $pp\pi$ & -0.81 \\
  \end{tabular}
\end{table}

\begin{table}[h!]
  \centering
  \caption{Onsites and cut-off lenghts.}
  \label{tab:table2}
  \begin{tabular}{ccccccc}
    \toprule

$\varepsilon_{s}(C)$&$\varepsilon_{p}(C)$&$\varepsilon_{s}(Ca)$&$\varepsilon_{p}(Ca)$&$r_{C-C}$&$r_{Ca-Ca}$&$r_{C-Ca}$  \\

    \midrule
$-13.6eV$&$-0.610eV$&$6.89$&$7.39eV$
&$1.48\AA$&$4.30\AA$&$2.52\AA$  \\

    \bottomrule
  \end{tabular}
\end{table}

\end{document}